\documentclass[12pt]{article}

\usepackage{graphicx}

\newcommand{\beq}{\begin{equation}}
\newcommand{\eeq}{\end{equation}}
\newcommand{\beqa}{\begin{eqnarray}}
\newcommand{\eeqa}{\end{eqnarray}}
\newcommand{\beqar}{\begin{eqnarray*}}
\newcommand{\eeqar}{\end{eqnarray*}}

\begin{document}
\thispagestyle{empty}

\hfill{\sc UG-FT-187/05}

\vspace*{-2mm}
\hfill{\sc CAFPE-57/05}

\vspace{32pt}
\begin{center}

\textbf{\Large 
A Little Higgs model of neutrino masses}
\vspace{40pt}

F.~del \'Aguila, M. Masip and J.L.~Padilla
\vspace{12pt}

\textit{
CAFPE and Departamento de F{\'\i}sica Te\'orica y del
Cosmos}\\ \textit{Universidad de Granada, E-18071, Granada, Spain}\\
\vspace{16pt}
\texttt{faguila@ugr.es, masip@ugr.es, jluispt@ugr.es}
\end{center}

\vspace{40pt}

\date{\today}

\begin{abstract}

Little Higgs models are formulated as effective theories with 
a cut-off of up to 100 times the electroweak scale. 
Neutrino masses are then a {\it puzzle}, since the usual 
{\it see-saw} mechanism involves a much higher scale that would 
introduce quadratic corrections to the Higgs mass parameter.
We propose a model that can {\it naturally} accommodate 
the observed neutrino masses and mixings in Little Higgs scenarios. 
Our framework does not involve any large scale or suppressed 
Yukawa couplings, and it implies the presence of three
extra (Dirac) neutrinos at the TeV scale.
The masses of the light neutrinos are induced radiatively, 
they are proportional to small ($\approx$ keV) mass parameters that 
break lepton number and are suppressed by the Little Higgs 
cut-off. 

\end{abstract}

\newpage

{\bf Introduction.}
The stability of the electroweak (EW) scale at the loop level
has been the main motivation to search for physics beyond the 
standard model (SM) during the past 30 years.
Namely, 
SM loops introduce quadratic corrections to the EW scale.
If this scale is {\it natural}, consistent with the dynamics 
and not the result of an
accidental cancellation between higher scales, then new physics
must compensate the SM quadratic contributions. 
In particular, top quark corrections to the Higgs mass squared
become of order $(500$ GeV$)^2$ for a cutoff $\approx 2$ TeV, 
what would suggest new contributions (from supersymmetry, 
extra dimensions,...) of the same order at the reach of the LHC.

Little Higgs (LH) ideas 
\cite{Georgi:1974yw,Arkani-Hamed:2001nc,Arkani-Hamed:2002qy}
provide another very interesting 
framework with a scalar sector
free of unsuppressed loop corrections. New symmetries
protect the EW scale and define consistent models with a 
cutoff as high as $\Lambda\approx 10$ TeV. Therefore, these models
could describe all the physics 
to be explored in the next generation of accelerators. 
More precisely, 
in LH models the scalar sector has a (tree-level) global 
symmetry G that is broken spontaneously at a scale
$f\approx 1$ TeV. The SM Higgses are then Goldstone 
bosons (GBs) of the broken symmetry, and 
remain massless and with a flat potential 
at that scale. Yukawa and gauge interactions break
explicitly the global symmetry. However, the models
are built in such a way that the loop diagrams giving non-symmetric
contributions must contain at least two different
couplings. This {\it collective} breaking keeps
the Higgs sector free of quadratic top-quark and
gauge contributions. At the same time, loops (and/or
explicit non-symmetric terms in the scalar potential)
give mass and quartic couplings to the Higgs, both
necessary to break the SM symmetry 
(see \cite{Schmaltz:2005ky} for a recent review).

The inclusion of neutrino masses in this framework 
looks problematic. In the SM these masses require a new scale
much larger than the EW one,
\beq
{\cal L}_{eff}= 
{1\over 2\Lambda_\nu} h^\dagger h^\dagger L L + {\rm h.c.}\;,
\label{eq1}
\eeq
where $h=(h^0\; h^-)$ and $L=(\nu\; e)$ 
are, respectively, the SM Higgs and lepton doublets.
At the EW phase transition $\langle h^0 \rangle = v$ 
and the neutrinos get their mass 
$m_\nu = v^2/ \Lambda_\nu\approx 0.1$ eV, 
what implies $\Lambda_\nu\approx 10^{14}$ GeV.
This effective (low-energy) scenario is simply realised 
using the {\it see-saw} mechanism \cite{seesaw}.
A SM singlet per family, $n^c$, is introduced 
with a large Majorana mass $M$ and 
sizeable Yukawa couplings $\lambda_\nu$ with the lepton 
doublets:
\beq
{\cal L}_\nu= \lambda_\nu h^\dagger L n^c + 
{1\over 2} M n^c n^c + {\rm h.c.}\;
\label{eq2}
\eeq
(the fields denote two-component left-handed spinors 
and family indices are omitted).
The spectrum is then three (Majorana) fields
of mass $\approx M$ plus the observed low-energy neutrinos 
with mass $m_\nu\approx \lambda_\nu^2 v^2/M$.

This scenario looks inconsistent in LH models, since the 
EW scale is not stable at the quantum level. 
\begin{figure}[b]
\centerline{
\includegraphics[width=0.3\linewidth]{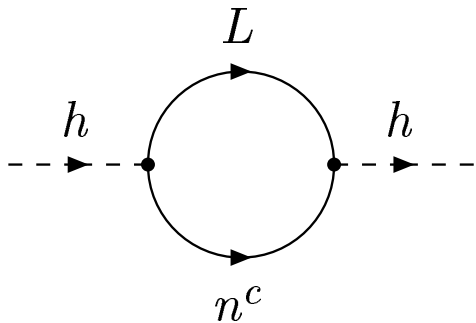}}
\caption{Diagram introducing corrections of order $M^2$ to
$m_h^2$. \label{fig1}}
\end{figure} 
In particular, the diagram in Fig.~1 gives a large 
contribution to the Higgs mass 
proportional to $M^2$:
\beq
\Delta m_h^2 \approx 
-{\lambda_\nu^2\over 8\pi^2} \left( 
C_{UV}+ {1\over 2} M^2 \left( 1- 2 \log{M^2} \right)
\right)\;,
\label{eq3}
\eeq
where we have used dimensional regularization and $C_{UV}$ 
contains the ultraviolet divergence
and renormalization-scale dependence. 
In principle LH models would avoid this type of corrections 
from heavy fields just because they are 
supposed to be effective theories valid only below 
a cutoff $\Lambda\approx (4\pi)^2 v$. However, 
if the {\it see-saw} scale $M$ is at (or below) the cut-off 
$\Lambda$, neutrino masses would be unacceptably large.
Therefore, the problem would be how to generate the operator in 
Eq.~(\ref{eq1}) without introducing also a term $\Delta m_h^2
h^\dagger h$ of order $M^2$ in the low-energy Lagrangian.
The option of not using the {\it see-saw} mechanism and assume
that neutrino masses are not different in origin from the masses
of quarks and leptons requires $M \approx 0$
and $\lambda_\nu\approx 10^{-12}$, a number that
would demand an explanation (see below). 
Other scenarios for neutrino masses 
\cite{Mohapatra:1991ng,Han:2005nk} which do not involve a
mass $M\gg 1$ TeV could be consistent with LH ideas. For
example, the model in \cite{Ma:2000cc} has $M\approx 1$ TeV
and extra scalars that give masses to neutrinos but not 
to quarks and charged leptons. 
The symmetries and structure of this
model (the scalars get VEVs of order MeV), however, 
seem difficult to accommodate in a simple LH model.
Here we propose a LH framework for neutrino masses 
that does not require large masses nor suppressed
Yukawa couplings.
It involves a quasi-Dirac field per family at the TeV 
scale and small ($\approx$ keV) Majorana mass 
terms breaking lepton number. 
The SM neutrinos are then Majorana fields that get their 
masses and mixings at the loop level.

{\bf Cancellations in Little Higgs models.}
LH models are able to explain why the top quark 
does not introduce one-loop 
quadratic corrections to $m_h^2$. In all the cases this
is achieved extending the global symmetries to the 
quark sector, so that the third generation appears in multiplets
of $SU(3)$. In particular, the doublet $Q=(t\; b)$ becomes a triplet
$\Psi_Q$.
In the same way, in order to build a consistent model 
of neutrino masses the global symmetry will be 
extended and the lepton doublets will become triplets. 

Let us focus on the {\it simplest} LH model 
\cite{Schmaltz:2004de,Contino:2003ve}, although our
arguments would be analogous in the original {\it littlest}
Higgs model \cite{Arkani-Hamed:2002qy} or other more 
complicated scenarios \cite{skibaterning,Kaplan:2003uc}. 
Here the scalar
sector contains two triplets, $\phi_1$ and $\phi_2$, of a global
$SU(3)_1\times SU(3)_2$ symmetry:
\beqa
\phi_1&\rightarrow & e^{i\theta^a_1 T^a} \phi_1\;,
\nonumber \\
\phi_2&\rightarrow & e^{i\theta^a_2 T^a} \phi_2\;,
\eeqa
where $T^a$ are the generators of $SU(3)$. To get gauge interactions,
the diagonal combination of the two $SU(3)$ is made local:
\beq
\phi_{1(2)}\rightarrow e^{i\theta^a(x) T^a} \phi_{1(2)}\;.
\eeq
At the scale $f$ the scalar triplets get vacuum expectation
values (VEVs) and break the global symmetry
to $SU(2)_1\times SU(2)_2$. For simplicity, 
it is usually assumed identical VEVs for both triplets
\beq
\langle \phi_{1}\rangle = 
\left(\begin{array}{c} 0 \\ 0 \\ f  \end{array}\right)\; , \;\;\;\;
\langle \phi_{2}\rangle = 
\left(\begin{array}{c} 0 \\ 0 \\ f  \end{array}\right)\;.
\eeq
The initial 12 scalar degrees of freedom in the two triplets 
contain 10 GBs plus two massive fields. 
However, these VEVs also break the gauge symmetry
$SU(3)\times U(1)_\chi$ to $SU(2)_L\times U(1)_Y$, a process that
will absorb 5 of the GBs. All this becomes apparent if 
the two triplets are parameterized
\beqa
\phi_{1(2)}=&&\!\!\!\! \exp\left\{{i\over f} 
\left(\begin{array}{cc} &h' \\ 
h'^\dagger & \eta' \end{array}\right) \right\} \times \nonumber \\ 
&&\!\!\!\!\exp\left\{ +\!(\!-\!)
{i\over f} \left(\begin{array}{cc}  &h  \\ 
h^\dagger & \eta \end{array}\right) \right\} 
\left(\begin{array}{c} 0 \\ f+{r_{1(2)}\over \sqrt{2}}  
\end{array}\right)\;,
\label{paramet}
\eeqa
where $h'$ and $h$ are (complex) $SU(2)$ doublets and $\eta'$, $\eta$,
$r_1$ and $r_2$ are (real) $SU(2)$ singlets. At the scale $f$ 
$h'$ and $\eta'$ are {\it eaten} by the massive vector bosons
of $SU(3)$, $r_{1,2}$ get massive, and $h$ (and possibly $\eta$) 
are the SM Higgses. Therefore
\beqa
\phi_{1(2)} &\approx& 
\left(\begin{array}{c} 0 \\ f+{r_{1(2)}\over \sqrt{2}}\end{array}\right) 
+\!\!(\!-\!)\;i(1+{r_{1(2)}\over f\sqrt{2}})
\left(\begin{array}{c} h \\ \eta \end{array}\right) \nonumber \\
&&-\;{1\over 2}(1+{r_{1(2)}\over f\sqrt{2}})
\left(\begin{array}{c} \eta h \\ h^\dagger h+\eta^2 \end{array}\right)\;.
\label{exp}
\eeqa

In this model the top-quark Yukawa sector includes a triplet
$\Psi_Q=(Q\; T)$ and two singlets ($t^c_1$,
$t^c_2$), and it is described by the Lagrangian
\beqa
{\cal L}_t &=& \lambda_1 \phi_1^\dagger \Psi_Q t_1^c + \lambda_2
\phi_2^\dagger \Psi_Q t_2^c + {\rm h.c.} \nonumber \\ 
&\supset & \lambda_t ( h^\dagger Q t^c +
f T T^c -{1\over 2f} h^\dagger h T T^c)
 + {\rm h.c.}\;,
\label{eq5}
\eeqa
where $t^c=(i/\sqrt{2})(t_2^c-t_1^c)$, 
$T^c=(1/\sqrt{2})(t_2^c+t_1^c)$, and we have taken 
$\lambda_1=\lambda_2=\lambda_t/\sqrt{2}$. 
The one-loop quadratic corrections in
Fig.~2 cancel, which reflects that if one of the $\lambda_{1,2}$ 
couplings is zero the global $SU(3)_1\times SU(3)_2$ 
symmetry would be exact in this sector.

\begin{figure}[t]
\centerline{
\includegraphics[width=0.7\linewidth]{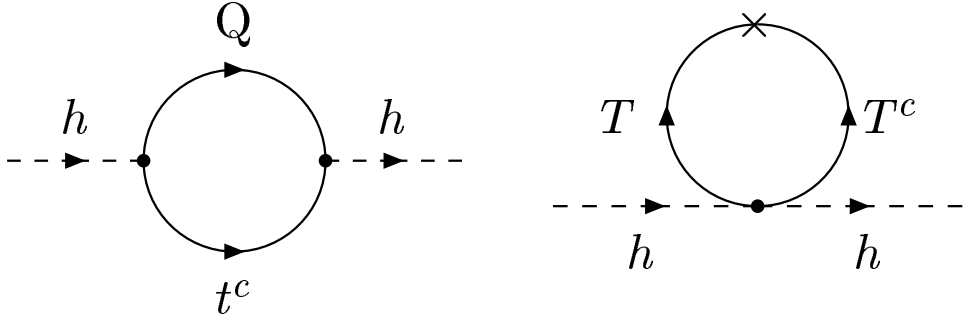}}
\caption{One-loop corrections to $m_h^2$ in LH models. \label{fig2}}
\end{figure}
Analogously, the model includes one lepton triplet 
$\Psi_L=(-iL\; N)$ (we use the $-i$ phase to simplify the
couplings) and one singlet $n^c$ 
per generation \cite{Lee:2005mb}. The Lagrangian
\beqa
{\cal L}_\nu &=& \lambda_\nu \phi_1^\dagger \Psi_L n^c + 
{\rm h.c.} \nonumber \\ 
&\supset & \lambda_\nu ( -h^\dagger L n^c +
f N n^c -{1\over 2f} h^\dagger h N n^c)
 + {\rm h.c.}
\label{eq6}
\eeqa
respects the global symmetries and does not generate 
one-loop quadratic corrections to $m^2_h$.
When the Higgs $h$ gets a VEV $v=174$ GeV and breaks the EW
symmetry, the Yukawa couplings induce mass terms
and define the matrix 
\beq
\begin{array}{cc} 
{\begin{array}{ccc}
\;\;\nu\;\;\;&\;\;\;\;\;\; N \;\;&
\;\;\;\;\;\; n^c\\
\end{array}}
& \\
\left(
{\begin{array}{ccc}
0 \;\;&\;\;0 \;\;& -\lambda_\nu v \\
0 \;\;&\;\;0 \;\;& \;\;\lambda_\nu f \\
-\lambda_\nu v \;\;&\;\;\lambda_\nu f \;\;& \;\; 0 \\
\end{array}}
\right)
&
{\begin{array}{c}
\nu \\
N \\
n^c \\
\end{array}} \\
\end{array}\;.
\eeq
The neutrino sector will then contain three Dirac
fields of mass $\approx \lambda_\nu f$ plus three 
massless neutrinos $\nu'\approx \nu + v/f\; N$. 
Several observations are here in order.

{\it (i)} The SM neutrinos are exactly massless at
this stage. To give them masses one would need extra
singlets (which could combine with them to define
Dirac neutrinos) and/or new terms that break lepton 
number (which could introduce Majorana masses for
the SM neutrinos). 

{\it (ii)} The neutrino sector in 
Eq.~(\ref{eq6}) does not break the $SU(3)_1\times SU(3)_2$ 
symmetry: both $\phi_1$ and $\Psi_L$ are triplets of 
$SU(3)_1$ and singlets of $SU(3)_2$. At the loop
level $\lambda_\nu$ will contribute to the (invariant)
operator $\phi_1^\dagger \phi_1$ in the scalar potential, 
but it will not introduce quadratic corrections to 
$m_h^2$ (which would require a symmetry-breaking 
operator $\phi_1^\dagger \phi_2$). Quadratic corrections to 
$m_h^2$ would appear at one loop if we add to the Lagrangian
a term $\lambda'_\nu \phi_2^\dagger \Psi_L n^c$,
and at higher order if we combine $\lambda_\nu$ with 
other Yukawa and gauge couplings.

{\it (iii)} Any realistic LH model needs a 
mass term $-m_h^2 \phi_1^\dagger \phi_2$ 
and a quartic coupling $\lambda (\phi_1^\dagger \phi_2)^2$ to
trigger EW symmetry breaking. These terms may appear 
at the loop level, from scalar couplings with the 
global symmetry-breaking 
sectors of the model \cite{Kaplan:2003uc}. Once these terms
are included, higher order corrections will induce the 
terms $\lambda'_\nu \phi_2^\dagger \Psi_L n^c$ 
in the Lagrangian.

{\bf A mechanism for neutrino masses in LH models.}
To obtain massive SM neutrinos we must then introduce
additional singlets or break lepton number. 
It is easy to see that the first possibility requires
very suppressed Yukawa couplings. An extra singlet 
(per generation) would make this sector identical to
the top-quark sector in Eq.~(\ref{eq5}), with two Dirac 
fields of masses proportional to $f$ and $v$.
The SM neutrinos would then be too heavy unless 
the corresponding Yukawa couplings are very 
suppressed. This escenario could be naturally realized 
in models with extra 
dimensions \cite{Arkani-Hamed:1998vp,Grossman:1999ra} 
or in holographic models \cite{Contino:2003ve}, 
where the Higgs appears as a composite particle of 
some strongly coupled dynamics.

Therefore, the scenario 
that we propose requires just one singlet $n^c$ per 
family and the breaking of lepton number.
The simplest way to parameterize this breaking is through
a Majorana mass for $n^c$, so we add the term 
${1\over 2} M n^c n^c + {\rm h.c.}$ in the Lagrangian in 
Eq.~(\ref{eq6}).
At lowest order the neutrino mass matrix would read
\beq
\begin{array}{cc} 
{\begin{array}{ccc}
\;\;\;\nu\;\;\;\;\;&\;\;\; N \;\;&
\;\;\;\;\; n^c\\
\end{array}}
& \\
\left(
{\begin{array}{ccc}
0 \;\;&\;\;0 \;\;& -\lambda_\nu v \\
0 \;\;&\;\;0 \;\;& \;\;\lambda_\nu f \\
-\lambda_\nu v \;\;&\;\;\lambda_\nu f \;\;& \;\; M \\
\end{array}}
\right)
&
{\begin{array}{c}
\nu \\
N \\
n^c \\
\end{array}} \\
\end{array}\;.
\eeq
\begin{figure}[b]
\centerline{
\includegraphics[width=0.6\linewidth]{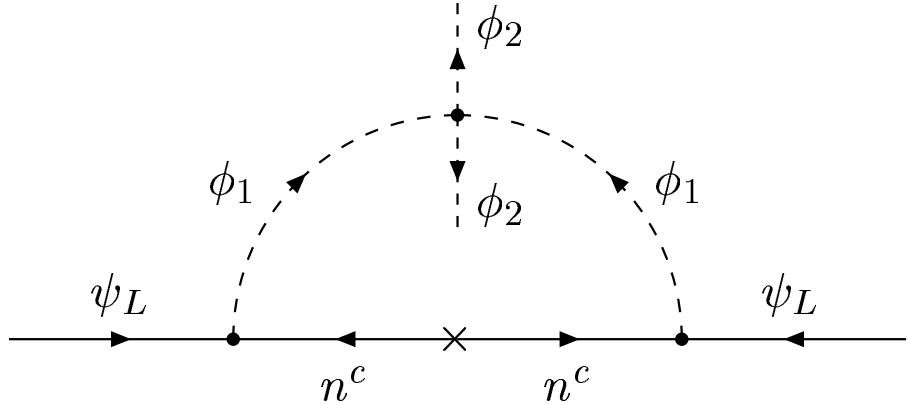}}
\caption{One-loop contribution to 
$(\phi_2^\dagger \Psi_L) (\phi_2^\dagger \Psi_L)$.
\label{fig3}}
\end{figure} 
This matrix implies two massive states and one massless 
neutrino per generation. The massless field, however, 
will get a mass at the loop level. The diagram
if Fig.~3 generates terms like 
\beqa
{\cal L}_{1} &=& {1\over 2\Lambda_\nu}
(\phi_2^\dagger \Psi_L) (\phi_2^\dagger \Psi_L)  + 
{\rm h.c.} \nonumber \\ 
&\supset & {1\over 2\Lambda_\nu} (h^{0\dagger} \nu +
f N ) (h^{0\dagger} \nu + f N ) + {\rm h.c.}
\label{final}
\eeqa
that will induce masses for the SM neutrinos. If $M < f$
and $m_h^2< (\lambda_\nu f)^2$ and assuming diagonal couplings
we obtain 
\beq
{1\over \Lambda_\nu}\approx { \lambda M \over 16\pi^2 f^2}
{x-1-\log x\over (x-1)^2}\;,
\label{inv}
\eeq
where $x\equiv m_h^2/(\lambda_\nu f)^2$ and $\lambda$ is the 
Higgs quartic coupling.
Notice that $1/\Lambda_\nu$ vanishes if 
$M=0$, since the term in Eq.~(\ref{final})
breaks lepton number. 
Combined with the terms
in Eq.~(\ref{eq6}) it also breaks the global
symmetries (it is proportional
to the Higgs quartic coupling $\lambda$). Although 
the mechanism 
to generate $\lambda$ may be different in each LH 
model, $\lambda$ must be 
$\approx 2m^2_h/v^2$.

The complete neutrino mass matrix is then 
\beq
\begin{array}{cc} 
{\begin{array}{ccc}\;\;\;\;
\nu\;\;\;\;\;\;&\;\;\;\;\;\; N \;\;&
\;\;\;\;\;\;\;\; n^c\\
\end{array}}
& \\
\left(
{\begin{array}{ccc}
\displaystyle { v^2/\Lambda_\nu} \;\;&
\;\;\displaystyle {vf/\Lambda_\nu} \;\;& 
-\lambda_\nu v \\
\displaystyle {vf/\Lambda_\nu} \;\;&
\;\;\displaystyle { f^2/\Lambda_\nu } \;\;
& \;\;\lambda_\nu f \\
-\lambda_\nu v \;\;&
\;\;\lambda_\nu f \;\;& \;\; M \\
\end{array}}
\right)
&
{\begin{array}{c}
\nu \\
N \\
n^c \\
\end{array}} \\
\end{array}\;.
\eeq
Its diagonalization gives two heavy neutrinos,
$n^c$ and $N'\approx N- v/f\; \nu$ (they define a 
quasi Dirac field), of mass  $\lambda_\nu f$ plus a SM neutrino 
$\nu'\approx \nu+ v/f\; N$ of mass (in the limit 
$m_h^2 \ll (\lambda_\nu f)^2$)
\beq
m_\nu\approx 4v^2/\Lambda_\nu\approx 
v^2{\lambda M \over 4\pi^2 f^2}
\log {(\lambda_\nu f)^2\over m_h^2}
\label{mass}
\eeq
per family.
To obtain $m_\nu\approx 0.1$ eV the Majorana mass $M$ must
be $\approx 0.1$ keV. 

A first interesting consequence of this result (that
distinguishes our framework from other scenarios for
neutrino masses) 
is that the light masses tend to depend only logarithmically 
on the Yukawa couplings.
Any difference or 
hierarchy in the Yukawa sector will appear {\it softened} by
the logarithm in the neutrino masses.

In our model the large mixings in the 
Maki-Nakagawa-Sakata (MNS) matrix \cite{Maki:1962mu} 
are obtained once 
the couplings $\lambda _\nu$ and the 
lepton number violating masses $M$ are allowed to be 
arbitrary $3 \times 3$ matrices. 
An acceptable rate for FCNC processes 
(mediated at one loop by the TeV singlets)
will require the alignment of the neutrino and the 
charged-lepton Yukawa couplings.
If the charged-lepton mass matrix comes from terms
$( \lambda _e / \Lambda ) \phi _1 \phi _2 \Psi _L e^c$ 
\cite{Schmaltz:2004de}, the rotations 
diagonalizing $\lambda _\nu$ and $\lambda _e$ must coincide 
to a large extent.
Then, the MNS matrix results from the diagonalization 
of Eq.~(\ref{mass}) (generalized to non-diagonal couplings) 
in the basis where $\lambda _\nu$ is diagonal.
Obviously, the allowed 
misalignment increases for larger values of $f$
(a detailed quantitative analysis will be presented 
elsewhere \cite{AMP}).

The model that we propose provides 
an explicit realization of TeV-scale non-decoupled 
neutrinos, which may be observable at a large ($e^+e^-$) collider 
\cite{delAguila:2005mf}. 
The fields $N$ have a mixing $\approx v/f$ with the SM 
neutrinos that must be smaller than 0.07 to 
be consistent with current data \cite{Bergmann:1998rg}. 
For $N$ masses of order TeV, 
single heavy neutrino production 
$e^+e^- \rightarrow N\nu \rightarrow eW\nu$ 
could give a signal at CLIC if  $v/f \ge 0.005$
\cite{delAguila:2005pf}.

{\bf Summary and discussion.}
In LH models the EW scale is protected from large quadratic
corrections only at the one-loop level. Therefore, the framework
can not naturally accommodate physics at scales larger 
than 10 TeV. Neutrino masses are then a puzzle, because in order to 
explain their size the effective low energy model must involve 
a much larger scale or very suppressed Yukawa couplings. 
To be consistent with a {\it see-saw}
mechanism of neutrino masses, LH models should incorporate 
{\it another} mechanism to suppress the SM
quadratic corrections at the ultraviolet 
cutoff $\Lambda\approx 10$ TeV. In that case the role of LH ideas
would be {\it just}, for example, to increase the scale of supersymmetry
breaking in one order of magnitude. On the other hand, the
possibility of very suppressed Yukawa couplings would imply
that the LH model is embedded in a theory with extra dimensions
or (its CFT dual) strongly coupled dynamics.

We have found a LH alternative that can
explain the small size of neutrino masses with no need
for a large scale nor extra dynamics. The lepton sector includes 
a gauge singlet $n^c$ per generation  
and has unsuppressed Yukawas, but it is free of one-loop 
quadratic corrections because of 
the same global symmetry as in the top-quark sector. This implies
another $SU(2)$ singlet $N$ per generation, with  
$n^c$ and $N'\approx N- v/f\; \nu$ combining into a 
massive field at the scale $f$ of global symmetry breaking.
As long as lepton number is not broken, the SM neutrinos
are massless. If $L$ is broken at a small scale 
$M\approx 10^{-6}$ GeV (a {\it mirror} scale of the one
in the {\it see-saw} mechanism) then the diagram in Fig.~3
introduces a term $\phi_2^\dagger \phi_2^\dagger
\Psi_L\Psi_L$ and the SM neutrinos get the
observed masses. 

In just the SM with an extra singlet per generation 
this mechanism would not work: the small value of 
the singlet mass $M$
would not prevent the SM neutrinos to combine with 
the singlets and define Dirac fields of mass 
$\lambda_\nu v$ similar to the mass of quarks and
charged leptons. However, 
the mechanism is naturally implemented in 
LH models, where the global symmetries imply new fields
and allow at the loop level the necessary couplings.

We would like to thank Ver\'onica Sanz for useful discussions.
This work has been supported by MCYT (FPA2003-09298-C02-01) and
Junta de Andaluc\'\i a (FQM-101). J.L.P.
acknowledges a FPU grant from MEC.

\end{document}